# BlockChain: A Distributed Solution to Automotive Security and Privacy

Ali Dorri, Marco Steger, Salil S. Kanhere, and Raja Jurdak

*Interconnected smart vehicles offer a range of sophisticated services that benefit the vehicle owners, transport authorities, car manufacturers and other service providers. This potentially exposes smart vehicles to a range of security and privacy threats such as location tracking or remote hijacking of the vehicle. In this article, we argue that BlockChain (BC), a disruptive technology that has found many applications from cryptocurrencies to smart contracts, is a potential solution to these challenges. We propose a BC-based architecture to protect the privacy of the users and to increase the security of the vehicular ecosystem. Wireless remote software updates and other emerging services such as dynamic vehicle insurance fees, are used to illustrate the efficacy of the proposed security architecture. We also qualitatively argue the resilience of the architecture against common security attacks.*

## Introduction

Smart vehicles are increasingly connected to roadside infrastructure, e.g. traffic management systems, to other vehicles in close proximity, and also more generally to the Internet, thus making vehicles part of the Internet of Things (IoT). This high degree of connectivity makes it particularly challenging to secure smart vehicles. Malicious entities can compromise a vehicle, which not only endangers the security of the vehicle but also the safety of the passengers. Miller and Valasek presented a sophisticated attack on a Jeep Cherokee using the wireless interface of the infotainment system whereby they were able to remotely control the core functions of the vehicle [1]. The data exchanged by the vehicle includes sensitive data, e.g., location, and can thus open up new privacy challenges.

Conventional security and privacy methods used in smart vehicles tend to be ineffective due to the following challenges:
- Centralization: Current smart vehicle architectures rely on centralised brokered communication models where all vehicles are identified, authenticated, authorised, and connected through central cloud servers. This model is unlikely to scale as large number of vehicles are connected. Additionally, the cloud servers will remain a bottleneck and a single point of failure that can disrupt the entire network.
- Lack of privacy: Most of the current secure communication architectures either do not consider user privacy, e.g. they resort to exchanging all data of the vehicle without the owner's permission, or reveal noisy or summarized data to the requester. However, in several smart vehicle applications, the requester needs precise vehicle data to provide personalised services.
- Safety threats: Smart vehicles have an increasing number of autonomous driving functions. A malfunction due to a security breach (e.g., by installing malicious SW) could lead to serious accidents thereby endangering the safety of the passengers and also of other road users in close proximity.

BlockChain (BC) is a distributed database that maintains a growing list of blocks that are chained to each other. BC was first proposed by Satoshi Nakamoto as the underlying technology behind Bitcoin [2]. BC has been shown to possess a number of salient features including security, immutability and privacy and could thus be a useful technology to address the aforementioned challenges. The structure of BC is shown in Figure 1. BC is managed distributedly by a peer to peer network. Each node is identified using a Public Key (PK). All communications between nodes, known as transactions, are encrypted using PKs and broadcast to the entire network. Every node can verify a transaction, by validating the signature of the transaction generator against their PK. This ensures that BC can achieve trustless consensus, meaning that an agreement between nodes can be achieved without a central trust broker, e.g. Certificate Authority (CA). A node will periodically collect multiple transactions from its pool of pending transactions to form a block, which is broadcast to the entire network. The block is appended to the local copy of the BC stored at a node if all constituent transactions are valid. A consensus algorithm such as Proof of Work (PoW), which involves solving a hard-to-solve easy-to-verify puzzle, is employed to control which nodes can participate in the BC. Once a block is appended, it (or the constituent transactions) cannot be modified, since the hash of each block is contained in the subsequent block in the chain, which ensures immutability. A node can change its PK (i.e. identity) after each transaction to ensure anonymity and privacy.

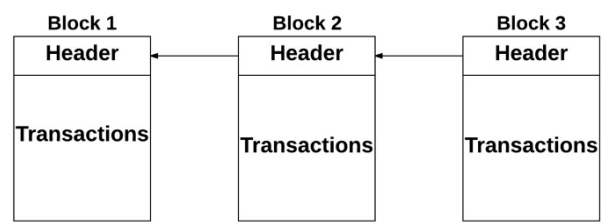

Figure 1. The structure of the BC.

BC has been used in a wide range of non-monetary applications, e.g. verifying proof of location [3]. In 2013, a new open-source BC-based platform, called Ethereum [4], was proposed to facilitate smart contracts, i.e. computer programs for enforcing a set of rules. BlockCharge [5] is a BC-based platform for electric vehicle charging. It uses Bitcoin as the underlying payment method and thus inherits the high-level of privacy offered by Bitcoin. The authors in [6] argued that the Ethereum BC can be used to distributedly create secure and private smart contracts between vehicle owners and service providers. However, a system that supports this is yet to be designed. In our previous work [7], we proposed an optimized BC instantiation for the Internet of Things (IoT) called Lightweight Scalable Blockchain (LSB).

The main contribution of this article is to present a decentralized privacy-preserving and secure BC-based architecture for the smart vehicle ecosystem. Smart vehicles, OEMs (i.e. car manufacturers) and other service providers jointly form an overlay network where they can communicate with each other. We base our design on LSB (a short overview is in the next section) due to its low overheads. Nodes in the overlay are clustered and only the Cluster Heads (CHs) are responsible for managing the BC and performing its core functions. These nodes are thus known as Overlay Block Managers (OBMs). Transactions are broadcast to and verified by the OBMs, thus eliminating the need for a central broker. To protect user privacy, each vehicle is equipped with an in-vehicle storage to store privacy sensitive data, e.g. location traces. The vehicle owner defines which data (and the granularity) is provided to third parties in trade for beneficial services and which data should be stored in the in-vehicle storage. Consequently, the owner has finer control over the exchanged data.

Vehicles can be mobile while communicating with the overlay. A vehicle that is physically distant from its associated OBM, may experience increased latency. To address this challenge, we propose to use a soft handover method (similar to Mobile IP [8]), wherein, the vehicle associates with a different OBM that is closer to its current location.

All transactions (i.e. communications) in the network are encrypted using asymmetric encryption. Nodes are authenticated using their PKs. Strong communication security and authentication introduced by BC mitigates the risk that the vehicle may be remotely hacked and thus increases the safety of the passengers.

## Overview of LSB

Conventional BC instantiations suffers from high (processing and packet) overhead and low scalability and throughput. The consensus algorithm employed in BC involves solving a hard-to-solve easy-to-verify puzzle that consumes significant computational resources. All transactions and blocks are broadcast to the entire network which results in pronounced packet overhead. Additionally, this raises a scalability issue as the number of broadcast packets increases quadratically with the number of participating nodes. The throughput of the BC is defined as the number of transactions that are stored in BC per second. Conventional BCs have limited throughput, e.g. Bitcoin throughput is restricted to seven transactions per second due to the complexity of the consensus algorithm. In our recent work [7], we developed a new BC instantiation, called LSB, that addresses the aforementioned challenges.

LSB is optimized for the IoT and large-scale low resource networks. LSB replaces the demand for solving a computational puzzle with a scheduled block generation process, thus eliminating the significant processing overhead of conventional BCs. Each node is permitted to store one block during a specific time period. To address the scalability challenge, LSB clusters the network and only the CHs (i.e. OBMs) manage the BC. LSB dynamically adjusts the throughput using a Distributed Throughput Management (DTM) method to ensure that the BC throughput does not significantly deviate from the transaction load generated by the nodes in the network. LSB uses a distributed trust algorithm to decrease the processing time associated with validating blocks. As shown in Figure 2, as more blocks are stored in BC, the processing time for validating new blocks in LSB is significantly lower compared to Bitcoin BC, which can be attributed to a novel distributed trust algorithm. In this article, we use LSB as the underlying BC technology motivated by its salient features outlined above.

Nodes use transactions to communicate with other nodes in the overlay. There are two types of transactions based on the

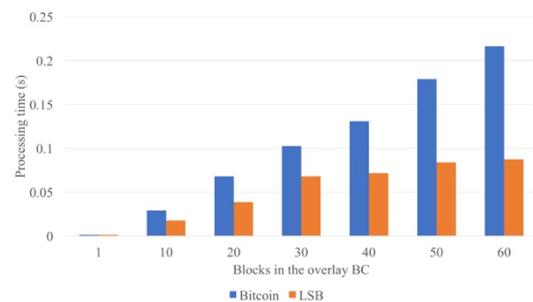

Figure 2. An evaluation of the processing time for validating new blocks [7].

number of signatures that must be validated:
1) *Single signature:* A single signature transaction requires one signature, which is the signature of the transaction generator, to be considered valid. The structure of this transaction is as follow:

$$\text{``T\_ID} \parallel P\_T\_ID \parallel PK \parallel Sig\text{''}$$

T_ID is the ID of the current transaction, which is the hash of the transaction. P_T_ID is the ID of the previous transaction of the transaction generator. It is used to link subsequent transactions of the same node, thereby creating a transaction ledger for that node. This is followed by the PK and signature (Sig) of the transaction generator.

2) *Multisig:* A multisig transaction requires two signatures, which are the signature of the transaction generator and recipient, to be considered valid. The structure of this transaction is as follows:

$$\text{``T\_ID} \parallel P\_T\_ID \parallel PK.1 \parallel Sig.1 \parallel PK.2 \parallel Sig.2\text{''}$$

T_ID and P_T_ID are the IDs of the current and previous transaction, respectively. The subsequent fields contain the PK and signature (Sig) of the transaction generator and recipient.

All transactions are broadcast to all OBMs. An OBM checks the validity of the received transaction by verifying the affixed signature(s). If the transaction is valid, then it is stored in a pool of valid transactions which will be collated to form a block with a pre-defined block size, i.e., the total number of transactions stored in the block. A multisig transaction that arrives at the OBM may yet need to signed by the recipient, particularly when the recipient belongs to the cluster of that OBM. Each OBM maintains a list of PK pairs (essentially an access control list) which establishes the nodes that are allowed to communicated with each other. The cluster members (i.e., overlay nodes) upload key pairs to the key list of their OBM to allow other overlay nodes to access them. If the OBM finds a PK pair in its list that matches with the PKs in the transaction (PK.1/PK.2), then it forwards the transaction to the corresponding node that uploaded the key pair. Otherwise, the transaction is broadcast to other OBMs. Figure 3 summarizes key LSB functions performed by an OBM.

## BlockChain-Based Architecture

In this section, we discuss the details of the proposed BC-based architecture for automotive security and privacy. The main part of our architecture is the overlay where a public BC is managed by the overlay nodes which can be smart vehicles, OEMs, vehicle assembly lines, SW providers, cloud storage providers, and mobile devices of users such as smartphones, laptops, or tablets. Figure 4 shows the overlay network.

Each vehicle is equipped with a Wireless Vehicle Interface (WVI) and local storage, such as a micro SD card. The WVI connects the vehicle to the overlay. The in-vehicle storage is used to store privacy sensitive data, e.g. location and maintenance history, to protect the privacy of the user. The vehicle generates single signature transactions in pre-defined time intervals containing the signed hash of the data stored in the in-vehicle storage. This transaction is sent to the OBM that the vehicle is associated with and thus stored in the BC. At a later time, the vehicle can prove that the data within its storage has not being changed by verifying the hash contained in this transaction. As the in-vehicle storage has limited capacity, a back-up storage can be considered in the smart home of the vehicle owner. The vehicle periodically transfers data from the in-vehicle storage to the backup storage. In this instance, the hash of the backup storage is stored in the BC.

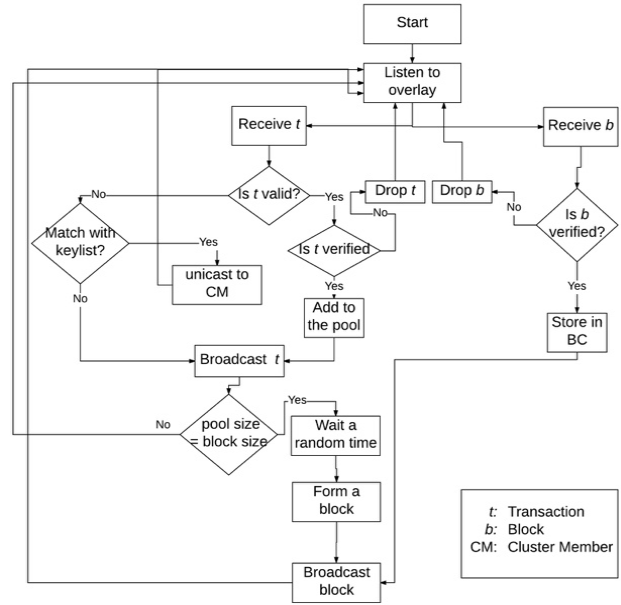

*Figure 3. A summary of the Overlay Block Manager (OBM) functions in LSB*

Overlay transactions are broadcast and verified by the OBMs. An OBM verifies a transaction by validating the signature of the transaction participants with their PK. Additionally, the OBM verifies if the previous transaction of each transaction, which is stored in the P_T_ID field, exists in the public BC.

Recall that, in BC each node is known by a changeable PK. Changing the PK for each transaction introduces a high level of privacy. However, in some instances other nodes may need to identify the real-world identity of a PK owner, e.g. the vehicles need to know the PK of their OEM so that they can trust requests sent from the OEM. To address this challenge, nodes whose identity should be known, including SW providers, OEMs, and cloud storage, share a PK that is certified by a third-party CA. Other overlay nodes can verify the CA's certificate to confirm the identity of these nodes. Note that, we rely on a centralized approach, i.e., existing public key infrastructure for this aspect of identity verification. However, the rest of the functionality is achieved by our proposed distributed architecture. It is worth noting that the aforementioned nodes can also use changeable PKs for transactions where their identity is to be kept private.

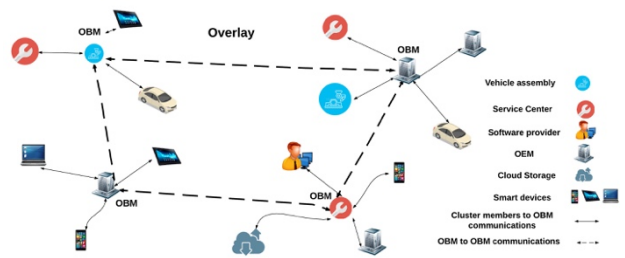

*Figure 4: An overview of the overlay.*

Recall that the overlay is clustered and cluster members use the OBM that they are associated with (i.e. CH), to send and receive transactions from the overlay. As vehicles move, they may experience extended delays in receiving responses from their OBM due to increased communication delays. We propose a solution that is based on the soft handover method [8]. When a vehicle moves to a new location, it measures the communication delay with multiple OBMs in its neighborhood. The OBM with the lowest delay is selected as the new OBM. Then, the vehicle updates the key list in this new OBM with a set of key pairs that allows other nodes to send transactions to this vehicle. Finally, the vehicle disconnects from the previous OBM, which clears the entries within its key list for the vehicle. Note that, as all transactions are broadcast to all OBMs, the new OBM will receive the transactions of the new vehicle that joined its cluster and thus this vehicle will continue to maintain connectivity with the rest of the overlay. In case the vehicle fails to find a suitable new OBM, e.g. if the OBMs are sparsely distributed, then the vehicle remains associated with the original OBM.

## Applications

In this section, we discuss various applications which can leverage the proposed architecture. Table 1 summarizes the key benefits of using BC compared to existing methods in each application which are discussed in more detail in the rest of the article.

## Remote Software Updates

The process of upgrading the functionality of the Electronic Control Units (ECU) of a vehicle or fixing a bug in the SW installed on one of the ECUs is known as Wireless Remote SW Update (WRSU). WRSU can be utilized during the vehicle development and assembly as well as for maintenance of the vehicle in a service center [9] or remotely from home. Securing WRSU is one of the most critical challenges in the automotive ecosystem, as it requires full access to the vehicle and its embedded control systems. Current security architectures are centralized, e.g. Tesla utilizes a VPN to perform remote software updates, which would not necessarily scale for very large number of smart vehicles. Furthermore, these architectures do not address the privacy issues outlined in the *Introduction*. Thus, WRSU demands a distributed security method while maintaining the vehicle owner's privacy.

The entire SW update process based on our architecture is sketched in Figure 5 and described in the following. Each OEM uses a cloud storage to store new SW updates so that its users can download the SW update. An account is created in the cloud storage for each vehicle by the OEM, and the account is associated with a public/private key pair. The keys are used to authorize and authenticate nodes who request to download the SW update.

First, the SW provider, which can be a specific department of the OEM or a supplier providing the ECU with the embedded SW, creates a new SW version and stores it in the cloud storage provided by the OEM (step 1 in Figure 5). Then, the SW provider creates a multisig transaction (see Section *A background on BC*) and populates its own PK in PK.1 field. The signed hash of the stored SW binary in the cloud is added to the Sig.1 field. As the binary is stored in the cloud, the hash can be verified by other overlay nodes thereby ensuring data integrity. Following this the SW provider populates the PK of the OEM in PK.2 field. Recall that OBMs use a key list to decide on how to forward a transaction. The SW provider sends the resulting multisig transaction to its OBM (step 2).

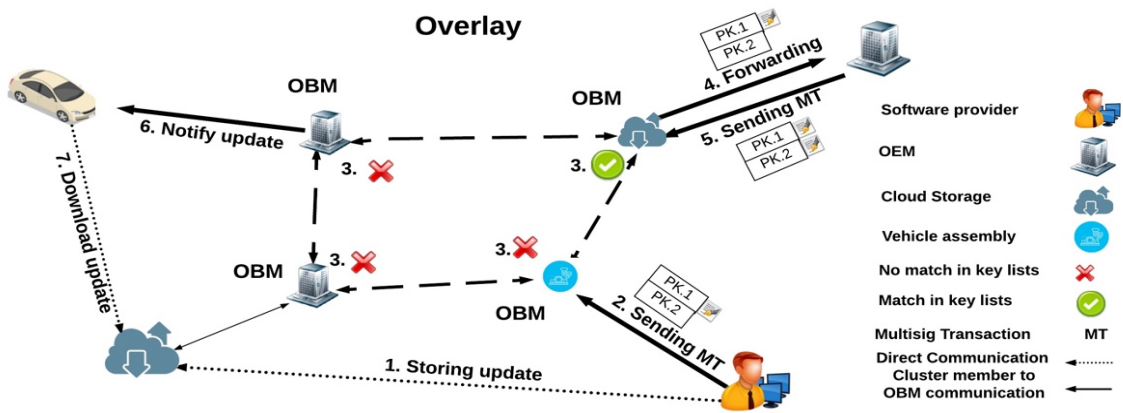

**Figure 5.** *WRSU process utilizing the BC architecture.*

Table 1. A summary of BC advantages compared to conventional methods employed for studied applications.

| Application | Conventional Methods | Advantages introduced by BC |
|---|---|---|
| WRSU | <ul><li>Centralized – not scalable</li><li>Partial participation: not addressing the full chain starting from a SP all the way to a service center</li><li>Lack of privacy: a direct link between the vehicle and OEM can compromise the driver's privacy (e.g., driver behavior or location)</li><li>Only OEM can verify communications or history of update downloads.</li></ul> | <ul><li>Distributed data exchange and security provides scalability</li><li>End-to-end: involving SP, OEMs, vehicles, service centers, assembly lines, etc.</li><li>Ensure privacy of the user (also for diagnostics)</li><li>Update history as well as authenticity of the SW can be publicly verified</li></ul> |
| Insurance | <ul><li>Current systems are often insecure, which endangers the vehicle's integrity [10]</li><li>Users lack control over the exchanged data</li><li>Privacy-sensitive data must be continuously sent to the insurance company for receiving services</li></ul> | <ul><li>Secure, distributed, and privacy-preserving data exchange</li><li>Users control the exchanged data</li><li>Privacy-sensitive data is shared on demand (e.g., accident happened) instead of a continuous data exchange. Authenticity of data stored in the vehicle can be publicly confirmed</li></ul> |
| Electric vehicles | <ul><li>Central payment and accounting</li><li>The location and behavior (e.g., using a specific charger on a specific day) of the user can be tracked.</li></ul> | <ul><li>Private and distributed security, payments and accounting</li><li>User data such as location information remain private</li></ul> |
| Car-sharing services | <ul><li>Central payment and accounting</li><li>Users can be tracked by their identity</li><li>Central authorization</li></ul> | <ul><li>Private and distributed security, payments and accounting</li><li>Users use changeable identities</li><li>Distributed authorization</li></ul> |

OBMs broadcast the transaction (step 3). The OBM of the cluster containing the concerned OEM, finds the match in its key list and thus forwards the transaction to the OEM (step 4). The OEM verifies the new SW version and signs the received transaction, by populating Sig.2 field. Then, the OEM sends the transaction to its OBM (step 5) which is then broadcast to all OBMs. The OBMs verify the multisig transaction by checking the signature of both the SW provider and the OEM using the PKs included in the transaction. Next, the OBMs notify their cluster members, i.e. vehicles, about the latest available SW update (step 6).

On receiving the transaction from the OBM, the smart vehicle verifies it by ensuring that the PK.2 filed in the transaction equates with the PK of its OEM. The vehicle subsequently downloads the SW directly from the cloud storage (step 7). Recall that each vehicle has a public/private key pair to authenticate itself to the cloud. Next, the vehicle verifies the integrity of the downloaded binary by comparing the signed hash of the SW binary in the received transaction, from the OEM and SW provider, with the hash of the downloaded version. This ensures the integrity during WRSU.

*Insurance*

Insurance companies are beginning to offer flexible vehicle insurance fees to their responsible customers. For this, the company evaluates the driving behavior using data collected from the vehicles such as braking patterns and speed. We now discuss the suitability of our architecture for this application.

Initially, when a car owner chooses such a flexible insurance model, the insurance company creates a public/private key pair for the car along with an account in a cloud storage. Thus, the insurance company knows the real identity of each account holder. The insurance company stores the PK in a secure database so that it can identify users later. The key pair is used by the vehicle for all subsequent communications (i.e. transactions) with the insurer. The vehicle stores data, e.g., braking pattern and speed, in the cloud storage using the provided account. This data is used by the insurance company to provide flexible insurance services to the user.

The insurance company knows the identity of the owner of the vehicle which stores data in the cloud storage. This endangers the vehicle owner privacy as the exchanged data might contain privacy sensitive data, e.g. the location of the vehicle. To address this challenge, such privacy sensitive data that might not necessarily be required for offering insurance services to the owner are stored in the in-vehicle storage. When this data is demanded by the insurance company, e.g. when an accident happens, the vehicle sends the data stored in the in-vehicle storage to the insurance company to file an accident claim. Recall that the hash of the in-vehicle storage is stored in BC.

This hash can be used by the insurance company to ensure that the data has not been modified since the time when the hash is stored in BC.

The vehicle owner may discontinue its contract with the insurance company or sell its vehicle. In such cases, the insurance company removes the vehicle account from the cloud storage, thus the vehicle is denied to receive further services and store data in the cloud storage.

*Electric Vehicles and Smart Charging Services*

The number of electric vehicles is constantly growing. This trend increases the demand for efficient and fast vehicle charging infrastructure. Interconnecting the smart vehicle to the owner's smart home and mobile devices, could lead to several sophisticated services. For example, the charging process can become more personalized, if information about the travel habits of the user are made available (e.g. through their calendar). This information can be used to guarantee that the vehicle is fully charged when the user needs it while also choosing the most efficient and cheapest charging cycle such as by avoiding peak load times.

The proposed security architecture allows the vehicle to exchange data with other IoT participants, e.g. smart home and smart devices of the user. These participants can be considered as the overlay nodes. The home (and vehicle) owner defines which information can be shared between these entities to protect his privacy while enabling novel services thus enriching the smart vehicle and its functionality. Blockcharge [5] can be used in conjunction with our method to pay the charging fees.

*Car-sharing Services*

Car sharing services, e.g. Car Next Door [11], are growing rapidly. Providing such highly distributed services requires the interconnection of smart vehicles, car-sharing service providers and the users of the services in a secure and reliable way. A trusted communication channel is needed to securely exchange data including the location of the vehicle, keys to unlock the car, and payment details of the user. The proposed security architecture is eminently suitable for these services as i) the decentralized nature of BC is tailor-made for these highly distributed services which include providing a user with the location of the car, handling the interconnection between the user and the car (i.e., to unlock and use the car) and payment/billing after using the car-sharing service, and ii) it interconnects the involved entities in a secure way while protecting the privacy of the users (e.g., no link between the real identity of an user and a certain route driven) and the vehicle from unauthorized access (i.e., only registered and authorized user are allowed to locate, unlock and use a vehicle).

**Security and Privacy Analysis**

In this section, we discuss the privacy and security of the proposed architecture.

*Privacy:* The privacy of the proposed method is inherited from the BC where each node uses a unique PK to communicate with other overlay nodes. This prevents malicious nodes from tracking an overlay node. Each vehicle is equipped with an in-vehicle storage to store privacy sensitive data. The vehicle owner can reveal data in the in-vehicle storage to the service providers in situations where this data is required (e.g. accident claim).

An attacker might attempt to deanonymize a user by linking different pieces of data associated with the same anonymous user (i.e., linking the PKs of the user). This attack, known as linking attack, endangers the privacy of the user. To protect against this attack, each user uses a fresh key for each of its interactions in the overlay.

*Security*: The security provided by our architecture can be largely attributed to the use of BC. Each transaction in BC contains the hash of the data which ensures integrity. All transactions are encrypted using asymmetric encryption methods which provide confidentiality. Recall that the OBMs maintain a key list that provides access control for cluster members in a way that only transactions for which the embedded PKs match with the key list in the OBM can be forwarded to a cluster member.

In the following, we evaluate the resilience of the proposed architecture to selected security attacks. Thereby, we focus on attacks affecting the security of a smart vehicle and define different attack scenarios that allow an attacker to take control of a vehicle:

**Changing a software binary in the cloud:** The attacker may seek to gain access to the cloud storage and manipulate the software binary with the goal of injecting malware to a large number of vehicles. In such instances, the hash of the infected binary differs from the hash included in the multisig transaction which is signed by the SW provider and the OEM. Thus, the vehicles can readily detect such an attack prior to installing the infected SW update.

**Distributing a false update by claiming to be the OEM or SW update provider:** The overlay nodes know the PK of the OEM and the SW provider. Therefore, the attacker cannot claim to be either of these entities as it requires the private key associated with the PK of the relevant entities.

**Distributed Denial of Service (DDoS) attack:** To orchestrate a DDoS attack, it is necessary to compromise a large number of vehicles in the overlay. The compromised vehicles send a large number of transactions to a targeted overlay node in order to overwhelm it. Recall that transactions are broadcast to all OBMs. An OBM forwards a transaction to a cluster member only if the keys in the transaction (i.e. PK.1 and PK.2) match with a key pair in the key list of the OBM. The overlay nodes authorize requesters to access them by uploading a key pair in the key list of the OBM. The transactions that are part of the DDoS attack would not generate a match in the key list and would thus be dropped and not impact the targeted node.

## Future Research Directions

In this section, we summarize future research directions:
- **Key management:** Each vehicle owns multiple keys for communication with SPs or users, which may change during the vehicle lifetime. Managing keys introduces a new research challenge.
- **Caching data:** Each connected vehicle, must download data, e.g. SW update, from a cloud which incurs packet overhead and delay in the overlay. Introducing caching in OBMs can reduce such overhead.
- **Applications:** The proposed architecture suits a broader range of applications, e.g. congestion control, that can be explored in more detail.
- **Mobility:** Frequent mobility of the vehicles increases the packet and processing overhead resulting from the handover process. New mobility-friendly methods can be introduced to reduce this overhead.

## Conclusion

In this article, we proposed a novel automotive security architecture based on BlockChain (BC). Due to its distributed nature, the proposed architecture eliminates the need for a centralized control and allows novel automotive services.

The privacy of the users is ensured by using changeable Public Keys (PK). The security of our architecture is largely inherited from the strong security properties of the underlying BC technology. Additionally, the OBMs provide access control for transactions sent to their cluster members. The architecture is able to support emerging automotive services by providing a secure and trustworthy way to exchange data while protecting the security of the end user.

We discussed several automotive use cases to illustrate the applicability of the proposed architecture. Additionally, we described possible attack scenarios and discuss how the proposed architecture is able to mitigate and inhibit these attacks.

## Biographies

Ali Dorri is a Ph.D. student in UNSW, Sydney, Australia. His research interests focus on enhancing security and privacy of the Internet of Things (IoT), smart vehicles, smart grids, energy management, smart cities, and healthcare. He is currently working on optimizing and adopting Blockchain for large-scale networks including IoT.

Marco Steger is senior researcher at VIRTUAL VEHICLE research center in Graz, Austria. He received a master's degree from Graz University of Technology in 2013 with a thesis titled "Development and Evaluation of C2X applications" done in cooperation with BMW AG, Munich, Germany. His research interests encompass dependable wireless and automotive communication networks, security in wireless/mobile/automotive networks, dependable wireless sensor networks, automotive software and control units, advanced driver assistance systems (ADAS), and vehicle-to-x communication.

Salil Kanhere received his PhD from Drexel University, USA. He is an Associate Professor in the School of Computer Science and Engineering at UNSW Sydney, Australia. His research interests include Internet of Things, pervasive computing, crowdsourcing, sensor networks and security. He has published 170 peer-reviewed articles and delivered over 20 tutorials and keynote talks. Salil is a Senior Member of both the IEEE and the ACM. He is a recipient of the Humboldt Research Fellowship.

Raja Jurdak is a Senior Principal Research Scientist at CSIRO, where he leads the Distributed Sensing Systems Group. He has a PhD in Information and Computer Science at UC Irvine. His current research interests focus on energy, mobility, and security in networks. Raja is an Honorary Professor at University of Queensland, and an Adjunct Professor at Macquarie University, James Cook University, and University of New South Wales. He is a Senior Member of the IEEE.